# The Mott criterion: So simple and yet so complex

Alexander Pergament, Genrikh Stefanovich, Nadezhda Markova

Petrozavodsk State University, Russia

E-mail: aperg@psu.karelia.ru

**Abstract**

September 30, 2015 marks the 110th anniversary of the birth of the famous English physicist N. F. Mott. This article is dedicated to his memory. Here we consider the problem of metal-insulator transition. It is shown that the Mott criterion $a_B(n_c)^{1/3} \approx 0.25$ is applicable not only to heavily doped semiconductors, but also to many other materials and systems, including strongly correlated transition-metal and rare-earth-metal compounds, such as vanadium oxides. Particular attention is given to a 'paramagnetic metal – antiferromagnetic insulator' transition in chromium-doped $V_2O_3$. In Supplement we also briefly consider the history and state of the art of the Mott transition problem.

1. Introduction

The Mott concept of a metal-insulator transition (MIT) appears to be amongst the most important 20[th]-century physical theories [1-7], along with, e.g., the BCS model for superconductivity, since it concerns a rather fascinating and interest-provoking problem of condensed matter physics. However, interestingly enough, the well-known Mott criterion describes "almost anything" but $V_2O_3$, whereas this vanadium oxide has ever been known to represent a *classic example of the Mott transition* [1-5].

The Mott criterion states that [1]

$$a_B n_c^{1/3} \approx 0.25, \tag{1}$$

where $a_B$ is the effective Bohr radius and $n_c$ is the critical carrier density for the transition to occur. This simple criterion provides a numerical prediction for a MIT in many different systems, from doped semiconductors and high-$T_c$ superconductors (HTSC) to metal-ammonia solutions and metal – noble gas alloys [4] (see Fig. 1 and references [4, 8-12]). As to the range of the Mott criterion's applicability, it is appropriate to quote a phrase from the preface to monograph [2]: "This book offers a collection of reviews on nonmetal-to-metal (metal-insulator or Mott) transitions in very different physical systems, from solids with a regular periodic structure via disordered fluids and plasmas, finite metal clusters to exotic nuclear and quark matter" (see Section 4: Supplement). What a wide field for reflections and applications, indeed! Nonetheless, the focus of the study in this very brief essay is only on MITs in various condensed-matter systems.

The Mott criterion expressed by Equation (1) proved to be very successful in describing MITs in various ordered systems, doped semiconductors included [1, 4, 11]. The Hubbard model suggests that "interactions between electrons are accounted for, via the repulsive Hubbard $U$ term, only when they are on the same site" [2] ; for disordered systems, Anderson showed that at a certain degree of disorder all the electrons would be localized, which renders the system non-conducting [1-4, 13].

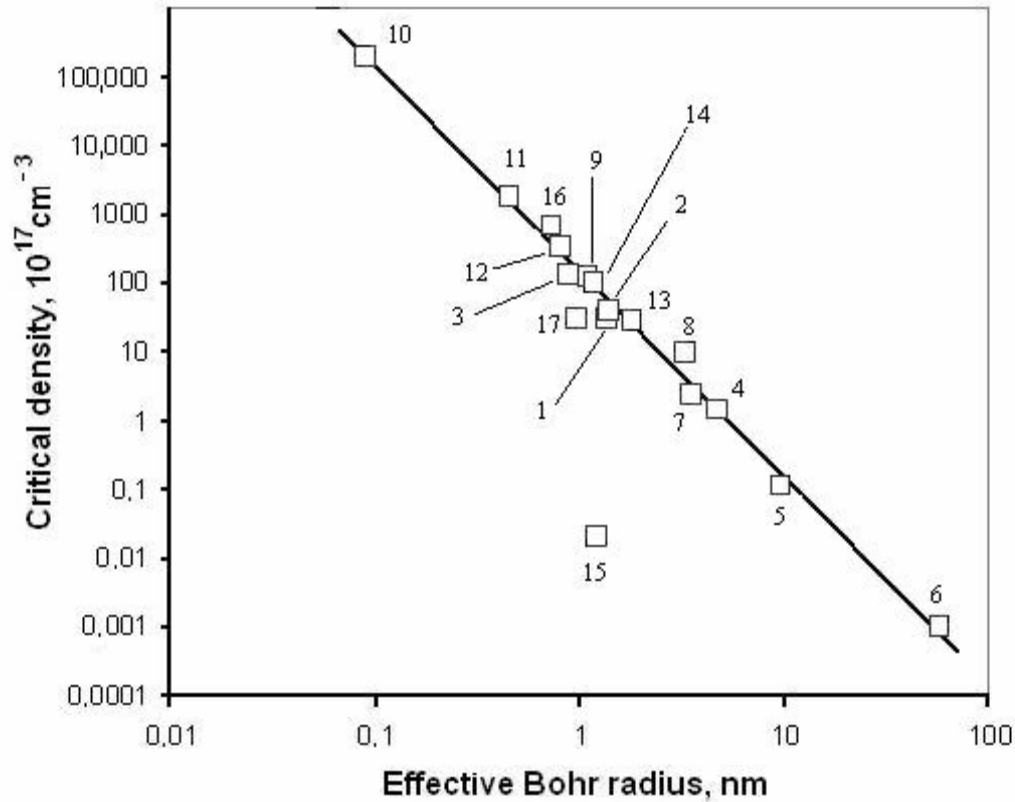

**Figure 1.** Correlation between the effective Bohr radii and critical densities for different systems[1]. 1 – Si:P, 2 – Si:B, 3 – Si:Bi, 4 – Ge:Sb, 5 – n-GaAs, 6 – n-InSb, 7 – CdS:In, 8 – CdS:Cl, 9 – GaP:Zn, 10 – Cu-Ar, 11 – YBa$_2$Cu$_3$O$_{7-\delta}$, 12 – NH$_3$-Na (see [11] for exact values of $a_B$ and $n_c$ and corresponding references), 13 – VO$_2$ [10], 14 – V$_2$O$_3$ (PI – PM), 15 – V$_2$O$_3$ (AFI – PM), 16 – SmS [8, 9, 11], 17 – Sr$_{1-x}$La$_x$TiO$_3$ [12]. Only some representative materials are included; other examples can be found in [4, 11]. Straight line corresponds to Equation (1).

As is known, there are two limiting descriptions of the outer electrons in solids, namely the band theory and the ligand-field theory [13]. In the elementary band theory, when atoms are combined to give a crystal with $N$ primitive unit cells, each atomic orbital (electron energy level) per unit cell gives rise to $N$ electron states in the energy band. Whether the material will be a metal or an insulator, depends on whether the band is filled partially or completely, or whether or not the two bands will intersect. The number of charge carriers can be increased by adding an impurity, and when the dopant concentration exceeds a certain value ($n_c$), a semiconductor undergoes a Mott MIT into a metal state. Similar processes are characteristic of other compounds of variable composition, such as, for instance, HTCS [13].

---

[1] It would be pertinent to remind here that such a vivid diagram was first represented, to all appearances, by Edwards and Sienko in 1978 (see references in [1; p.151] and [4]).

In compounds exhibiting a temperature- or pressure-induced MIT, instead of the one caused by doping or substitution, the Mott transition theory is more sophisticated [1-9]. As compared with the above simple picture, some new important phenomena are observed in compounds of $d$ and $f$ elements. For example, transition metal compounds containing atoms with unfilled $d$-shells form complex systems of phases with multiple oxidation states and mixed valence. These compounds belong to a class of strongly correlated systems, and strong electron-electron correlations are associated with a specific behavior of $d$-electrons. Since the $d$-bands are narrow, the energy of electron-electron Coulomb interactions is of the order of the bandwidth $B$ (and, hence, the electron kinetic energy), i.e., $U \sim B$. It is commonly accepted by now that strong correlation effects are responsible for some unique properties of such materials as HTCS, materials with metal-insulator transitions, heavy-fermion superconductors, ferromagnetic perovskites with colossal magnetoresistance, etc.

Another well-known example of the influence of correlation effects on the band structure is a "correlated insulator" (NiO, for example). Insulators of this type are also known as Mott insulators, for it was Mott's suggestion that electron repulsion is responsible for a breakdown of normal band properties for the $d$ electrons, which was later formalized in the Hubbard model [1, 3, 13]. Note that high- and low-temperature phases of vanadium sesquioxide doped by chromium to *ca* 1 at. % are just the ones representing, amongst others, such Mott insulator phases [1, 5, 14].

2. Results and discussion

The low-temperature AFI semiconducting phase resistivity of $V_2O_3$ close to the AFT – PM transition is $\rho \sim 5 \times 10^3$ $\Omega \cdot cm$ [15]. Knowing the effective charge carrier mass $m^* = 9m_e$, their mobility $\mu = 0.6$ $cm^2V^{-1}s^{-1}$ [1, 11], and dielectric permittivity of the material $\varepsilon = 8.5$ [16], it is straightforward to calculate the product on the left-hand side of Equation (1), which yields $6.3 \times 10^{-4} \ll 0.25$ [11]. We thus observe a significant divergence from Equation (1) for the AFI – PM transition. A similar analysis of a high-temperature transition is given in [11] where Equation (1) for the PI – PM transition in $V_2O_3$:Cr is shown to be valid.

The point is that there exists a mixture of metallic and dielectric phases close to the MIT in $V_2O_3$ [5]. The properties of such mixed regions are generally modeled in terms of the percolation theory and the behavior of the system is governed by its closeness to the percolation threshold [15]. In this case, one obtains the effective Bohr radius for charge carriers:

$$a_B = (0.53 \text{ Å}) \frac{\varepsilon}{m^*/m_e} = 1.18 \text{ nm}, \qquad (2)$$

where 0.53 Å is the hydrogen atom (or, genuine Bohr) radius and $\varepsilon = 200$ [11, 17] is the dielectric permittivity of the $V_2O_3$ insulating phase in immediate proximity to the transition. The ensuing $a_B$ value of ~ 1.2 nm seems to be quite a reasonable estimate and, moreover, it conforms to the values of $a_B$ and the MIT correlation length $\xi$ in another vanadium oxide, namely, $VO_2$ [10, 18]. The PI-phase resistivity just at the PI – PM transition onset point is $\rho \sim 1$ Ω·cm [15] and the value of the charge carrier mobility is again assumed to be $\mu = 0.6$ cm$^2$V$^{-1}$s$^{-1}$. Calculating the critical concentration from these values gives $n_c \approx 10^{19}$ cm$^{-3}$. Now we obtain $(n_c)^{1/3}a_B = 0.254$ (point 14 in Fig. 1), which is in full agreement with the Mott criterion. This suggests that for the PI → PM transition in $V_2O_3$ the Mott's law can truly be valid [11].

In the case of low-temperature AFI → PM transition, the situation becomes more complicated. There, as indicated above, the product of $a_B$ and $n^{1/3}$ is far less than 0.25, i.e., the Debye screening length is far in excess of $a_B$, which suggests that the Mott transition to a metallic state would seem hardly probable. If we consider a classic case, the screening length is expressed as

$$L_{D,\text{class}} = \sqrt{\frac{k_B T \varepsilon_0}{4\pi e^2 n}}, \qquad (3)$$

and we still obtain $L_D > a_B$: for $T = T_t = 170$ K, Equation (3) yields $L_D = 5.6$ nm, while $a_B = 1.2$ nm, see Equation (2).

Generally, expressions of the type $L_D = R$ (where $R$ is the electron localization radius, for instance, $a_B$) or $B = U$ (where $B$ and $U$ are the kinetic and potential electron energies, respectively) are nearly equivalent to the Mott criterion given by Equation (1): see, e.g., discussions in [19, 20] related to this issue. Unfortunately, we do not know the exact values of $\varepsilon$ and $m^*$ near the phase transition because of their divergence (or unrestricted growth), and probably that very factor is responsible for the fulfilment of the equality $L_D = R$. This situation is illustrated in Fig. 2 which suggests that an intermediate phase is present and that an increase of $R = a_B$ due to increased $\varepsilon$ close to the MIT, as well as a decrease of $L_D$ due to increased $n$, would result in the intersection of the two curves at the transition point.

It should be noted that, until recently, it was generally agreed that in most cases the Mott criterion did not apply to a MIT in transition metal compounds. Such an argument is based on very large values of $\varepsilon$ needed for the model of hydrogen-like impurities to be applicable [13]. However the divergence (or at least a considerable increase) of $\varepsilon$ close to the MIT renders the above restriction invalid (Fig.2, a). Hence, the statement made in [13], namely, that "the Mott criterion has been found to apply quite well, *outside the oxide field*, to transitions occurring over an enormous range of $n_c$ [4]", appears too categorical. The primary emphasis in this study is just on the postulate that transition metal compounds exhibiting a MIT, particularly oxides, can also be included in the general picture described by the Mott criterion. This general picture is represented by the correlation observed in Fig. 1. The only point 15 for the AFI – PM transition in $V_2O_3$:Cr is outside the straight line, corresponding to the constant 0.25 on the right-hand side of Equation (1). The reason for such an inconsistency is discussed above.

Finally, we would like to mention a rather curious, even amusing, fact. In recent years, a number of papers have appeared concerning a "metal-insulator transition" in $V_2O_5$ [22, 23]. We do not presume to dispute the essence of their matter. The case in point will be different.

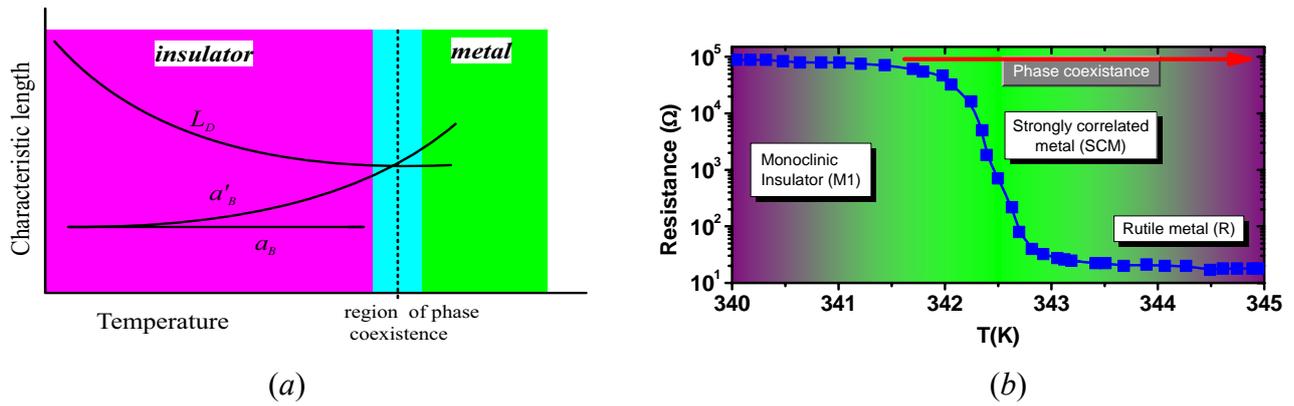

**Figure 2**. Intermediate state in vanadium dioxide (*a*) treated as a strongly correlated metal (*b*) [7]. Panel (*a*) shows schematic temperature dependences of the Debye screening length and localization radius. If account is taken of the dependence of $a_B$ on $T$ via the divergence of $\varepsilon$ near the transition point (curve $a'_B$), the transition does occur (vertical dashed line). For more detailed discussion of the $L_D$ dependence on $T$, see, for example, references [7, 9]. A region of phase coexistence (termed as an "intermediate state" [10, 18] or a "strongly correlated metal" [21], panel (*b*), in case of $VO_2$) with a divergent effective mass is also shown.

The fact is that the term "metal-insulator transition", particularly with respect to vanadium oxides (and not only to these – generally to a variety of transition metal oxides and related materials, as is shown above), has long been approved in the scientific literature to describe quite a certain range of phenomena.

The above-described 'harmonious' picture exploded when we learned about a "MIT" in vanadium pentoxide [22, 23] which is known to be a semiconductor with the band gap width of ~ 2.2 eV [13] and exhibits no trend to a transition into metallic state. Here the point is that we are dealing with a change in phase composition, which is not the same thing, we should say. We are presented with a phase transition of vanadium pentoxide upon heating up to ~ 280°C [23] to *other* vanadium oxide phases, exhibiting metallic properties at these temperatures, as purportedly does a MIT in $V_2O_5$. It's like taking a piece of aluminum, forming an anodic oxide on it, and saying afterwards that, as a result of anodizing, a metal-to-insulator transition occurred.

The above story originates from the paper [24] where experiments on memory switching in thin-film $V_2O_5$-based structures were interpreted in terms of the idea of a "glass → metal" transition in amorphous vanadium pentoxide. This idea was further widely circulated (with no efforts to verify the experimental results) and much referred to. The reference was first made in [25] (just as with a MIT in $V_2O_5$ at $T_t = 257$ °C), followed by a few refutations [26, 27]. The subject could have seemed exhausted. Alas, history repeats itself in the 21st century. In the papers [18, 28, 29] (see also additional references concerning this matter in [30, 31]) we again encountered the same statement – and again with a reference to that old paper [24] – and, finally, apotheosis: such respectable journals as Physical Review Letters and Applied Physics Letters publish papers on a "MIT in $V_2O_5$" as it is asserted in the titles of the references [22, 23]. We should say that all these issues regarding the alleged "MIT in $V_2O_5$" have recently been discussed in a number of our articles and comments (see, e.g., the works [30-32] as well as our review [7]), where all this is analyzed more or less in detail.

It should be admitted, however, that everything is honest in those papers (we mean [22, 23]), but entering such terms in the title looks like a desire to earn dividends on an inflated sensation. There is no, and cannot be, a metal-insulator transition in vanadium pentoxide in the commonly-accepted sense discussed above. Even if one tries to put a corresponding point in Fig.1, it certainly will exceed the bounds of the figure, well below the abscissa axis, solely because of too low carrier density in dielectric $V_2O_5$. And the last thing to do: the scientific importance of the works [20, 21] is beyond question, we would just ask not to use the "metal-insulator transition" term, for this is confusing in such a context as we've tried to show above.

3. Conclusion

Frankly speaking, the primary goal of this paper was to somehow elucidate this knotty question, namely, why the transition metal compounds, $V_2O_3$ in the first place, do not always obey the Mott criterion (and we are once again reminded that $V_2O_3$ represents a classic example of a material exhibiting the Mott transitions). One cannot but admit that the present attempt seems to fail to clarify the situation: the AFI – PM transition (point 15 in Fig.1) does not obey this criterion, and Fig. 2 only qualitatively shows how this discrepancy could be avoided. Nevertheless, we believe that these notes could contribute to further understanding of the subject matter of our

study – the Mott criterion – which turned out to be so simple and yet so complex. After all, one always needs very simple things, although a more complex subject is often better understood: "Men quickly grasp the complex schemes, When simplicity's their greater need…" [33].

It should be emphasized that the model discussed by no means offers the exact MIT mechanism which is more complex and includes more subtle effects in transition-metal (*d*) and rare-earth-metal (*f*) compounds, compared to doped semiconductors. The model describing MITs in those compounds on the basis of the Mott criterion is not supposed to predict or explain any specific properties or experimental data. The only aim of this work is to show that the concept of the Mott criterion is universal and must work in all the materials undergoing the Mott transition, regardless of whether it is a simple Si:P or a complex $V_2O_3$:Cr. It seems quite appropriate here to draw the analogy to the situation with the theory of superconductivity, where the Cooper pairing model is common for different mechanisms, be it electron-phonon interaction, bipolaron mechanism, RVB magnon-based mechanism, etc.

Summarizing, it is shown in the present study that the Mott criterion is applicable not only to heavily doped semiconductors, but also to many other materials, including such important for the MIT problem on the whole as $VO_2$, SmS (see the caption to Fig.1) and, especially, $V_2O_3$. In order to verify model approximations introduced in this article, additional experimental data are likely to be needed. Particularly, further studies seem to be of importance, focused on thorough measurements of the dielectric constant, effective mass and carrier mobility values of both vanadium oxide Magneli phases [30, 34] and other strongly correlated *f*- and *d*-compounds undergoing Mott metal-insulator transitions.

4. Supplement

4.1. Brief history excursus

This subsection represents a compilation mainly based on an article [35] in Wikipedia. Although the band theory of solids had been very successful in describing various electrical

**9**

properties of materials, in 1937 J. H. de Boer and E. J. W. Verwey pointed out that a variety of transition metal oxides predicted to be conductors by band theory (because they have an odd number of electrons per unit cell) are insulators [36]. Later on (also in 1937) N. F. Mott and R. Peierls predicted that this anomaly can be explained by including interactions between electrons [37].[2)]

In 1949, in particular, Mott proposed a model for NiO as an insulator, where conduction is based on the formula [38]

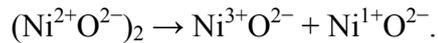
$$(Ni^{2+}O^{2-})_2 \rightarrow Ni^{3+}O^{2-} + Ni^{1+}O^{2-}.$$

In this situation, the formation of an energy gap preventing conduction can be understood as the competition between the Coulomb potential $U$ between $3d$ electrons and the transfer integral $t$ of $3d$ electrons between neighboring atoms (the transfer integral is a part of the tight-binding approximation). The total energy gap is then

$$E_{gap} = U - 2zt,$$

where $z$ is the number of nearest-neighbor atoms.

In general, Mott insulators occur when the repulsive Coulomb potential $U$ is large enough to create an energy gap. One of the simplest theories of Mott insulators is the 1963 Hubbard model. The crossover from a metal to a Mott insulator as $U$ is increased can be predicted within the so-called dynamical mean field theory.

Apparently, it is in this paper of 1949 [38] that the Mott criterion was first introduced; what occurred later was only a clarification of the right-hand side constant in equation (1). Information concerning these facts can be found in monograph [1; page 2] (see Figure 3(*b*)). Finally, it should be noted that this year, 2014, marks exactly 65 years since the appearance of the Mott criterion (1), and this historic date, along with the next-year 110th anniversary of Sir N. F. Mott's birth, makes all the aforesaid especially important and even symbolical (Figure 3).

---

[2)] The surprising and, perhaps, a unique case – Mott and Peierls write a joint article. It is as if for example, Wagner and Verdi peacefully would play duets on the piano. If only they had known how the fate would "separate" them (not in life, of course, but in the legend) after MIT in vanadium dioxide was discovered and came to be studied intensively in the 70s–80s. All this will result in a real 'war' between the followers of the Mott transition or Peierls transition in $VO_2$. But that's another story ...

The question was, at what value of $a$ would a metal–insulator transition occur? The assumption (Mott 1949) was made that this would occur when the screened potential round each positive charge,

$$V(r) = -\frac{e^2}{r} e^{-qr}, \qquad (1)$$

with the screening constant $q$ calculated by the Thomas–Fermi method, was just strong enough to trap an electron. The transition would be discontinuous; with varying $n$, and thus $q$, there would be a discontinuous transition from the state with all electrons trapped to that where all are free. It was found that this would occur when

$$n^{1/3} a_H = 0.2 \qquad (2)$$

where $n$ is the number of electrons per unit volume and $a_H$ the hydrogen radius. Equation (2) was applied, with success, to the metal–insulator transition in doped semiconductors, the disorder resulting from the random positions of the centres being neglected. The disorder was believed to be the cause of the absence of a discontinuous change in the conductivity. However, the success of (2) was fortuitous; if the many-valley nature of the conduction bands of silicon and germanium is taken into account, and also the central cell corrections, totally different results are obtained (Kreiger and Nightingale 1971, Martino 1973, Green *et al.* 1977).

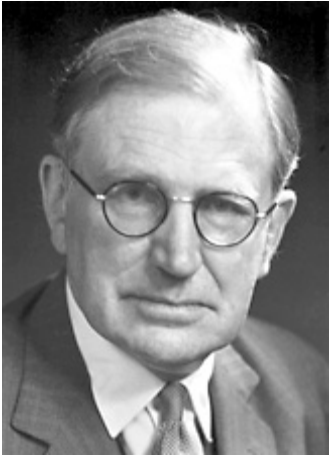

(*a*)        (*b*)

**Figure 3.** (*a*) Sir Nevill Francis Mott (1905-1996). Nobel Prize in Physics (1977). Mott criterion (1949 [38]). (*b*) Scan of the page 2 from [1].

4.2. State of the art

As we have seen, at present, the Mott transition concept is used quite widely in various fields of physics - from condensed matter physics to plasma physics [2]. In condensed matter physics, the Mott criterion works not only in doped semiconductors, but in many transition metal oxides as well. Furthermore, researchers consider, for instance, ultracold fermions in a one-dimensional bipartite optical lattice and metal-insulator transitions driven by shaking in such a system [39]. We remember next that, though some researchers have already relatively long since speculated about a possible quantum Mott-like phase transition in quark-gluon plasma [2], only recently long-wavelength Higgs modes in a neutral, two-dimensional superfluid close to the quantum phase transition into a Mott insulating state have been identified and studied [6]. In conclusion, we have to mention another recent discovery, namely, direct observation of the Higgs mode in a superconductor NbN excited by intense electric field transients [40]. The latter two facts perhaps once again do prove an intimate link between the Mott transition and superconductivity, do not they? Let's wait and see.

**Acknowledgements**

This work was supported by the Strategic Development Program of Petrozavodsk State University (2012 – 2016) and the RF Ministry of Education and Science as a part of state program in the scientific field, projects no. 2014/154 (#1704) and no. 3.757.2014/K. The authors are also grateful to Ms. V. Novikova and Dr. D. Kirienko for their help in the manuscript preparation.

**References**

[1] N. F. Mott, Metal-Insulator Transitions, Taylor and Francis, London, UK, 1990.

[2] R. Redmer, F. Hensel, and B. Holst, Eds., Metal-To-Nonmetal Transitions, vol. 132 of Springer Series in Materials Science, Springer, New York, NY, USA, 2010.

[3] M. Imada, A. Fujimori A, and Y. Tokura, "Metal-insulator transitions," *Reviews of Modern Physics,* vol. 70, no. 4, pp. 1039-1263, 1998.

[4] P. P. Edwards, T. V. Ramakrishnan, and C. N. Rao, "The Metal-Nonmetal Transition: A Global Perspective," *J. Phys. Chem.*, vol. 99, no. 15, pp. 5228-5239, 1995.

[5] S. Lupi, L., B. Mansart, A. Perucchi, A. Barinov, P. Dudin, E. Papalazarou, F. Rodolakis, J.-P. Rueff, J Baldassarre.-P. Itié, S. Ravy, and D. Nicoletti, P Postorino, P Hansmann, N Parragh, A Toschi, T Saha-Dasgupta, OK Andersen, G Sangiovanni, K Held, and M Marsi "A microscopic view on the Mott transition in chromium-doped $V_2O_3$," *Nature Communications*, vol. 1, no. 8 (105), doi:10.1038/ncomms1109, 2010.

[6] M. Endres, T. Fukuhara, D. Pekker, M. Cheneau, P. Schauβ, C. Gross, E. Demler, S. Kuhr, and I. Bloch, "The 'Higgs' amplitude mode at the two-dimensional superfluid/Mott insulator transition," *Nature*, vol. 487, pp.454–458, 2012.

[7] E. L. Kazakova, O. Y. Berezina, D. A. Kirienko, and N. P. Markova, "Vanadium Oxide Gel Films: Optical and Electrical Properties, Internal Electrochromism and Effect of Doping," *Journal on Selected Topics in Nano Electronics and Computing*, vol.2, pp. 7-19, 2014.

[8] V. Železný, J. Petzelt, V. V. Kaminski, M. V. Romanova, and A. V. Golubkov, "Far infrared conductivity and dielectric response of semiconducting SmS," *Solid State Communications*, vol. 72, no. 1, pp. 43–47, 1989.

[9] V. V. Kaminskii, L. N. Vasil'ev, M. V. Romanova, and S. M. Solov'ev, "The mechanism of the appearance of an electromotive force on heating of SmS single crystals," *Physics of the Solid State*, vol. 43, pp. 1030–1032, 2001.

[10] A. Pergament, "Metal insulator transition: the Mott criterion and coherence length," *J. Phys.: Condens. Matter*, vol. 15, no. 19, pp. 3217-3223, 2003.

[11] A. Pergament and G. Stefanovich, "Insulator-to-metal transition in vanadium sesquioxide: does the Mott criterion work in this case?" *Phase Transitions*, vol. 85, no. 3, pp. 185–194, 2012.

[12] S. Hui, Evaluation of Yttrium-doped $SrTiO_3$ as a solid oxide fuel cell anode [Ph.D. thesis], 2000, Available at: http://digitalcommons.mcmaster.ca/opendissertations/1659.

[13] P. A. Cox, Transition Metal Oxides: An Introduction to Their Electronic Structure and Properties, Clarendon Press, Oxford, UK, 1992.

[14] G. Keller, K. Held, V. Eyert, D. Vollhardt, and V. I. Anisimov, "Electronic structure of paramagnetic $V_2O_3$ : Strongly correlated metallic and Mott insulating phase," *Phys. Rev.*, vol. 70, no. 20, Article ID 195106, 2004.

[15] H. Kuwamoto, J. M. Honig, and J. Appel, "Electrical properties of the $(V_{1-x}Cr_x)_2O_3$ system," *Phys. Rev. B*, vol. 22, no. 6, pp. 2626-2636, 1980.

[16] H. Abe, M. Terauchi, M. Tanaka, and S. Shin, "Electron Energy-Loss Spectroscopy Study of the Metal-Insulator Transition in $V_2O_3$," *Jpn. J. Appl. Phys.*, vol. 37, no. part 1, no, pp. 584-588, 1998.

[17] H. Jhans, J. M. Honig, F. A. Chudnovskiy, and V. N. Andreev, "AC Conductivity in the Antiferromagnetic Insulating Phase of the $V_2O_3$ System," *J. Solid State Chem.*, vol. 159, no. 1, pp. 41-45, 2001.

[18] J. Nag, The solid-solid phase transition in vanadium dioxide thin films: synthesis, physics and application [Ph.D. thesis], 2011, Available at: http://etd.library.vanderbilt.edu/available/etd-04202011-182358/.

[19] J. Spałek, J. Kurzyk, R. Podsiadły, and W. Wójcik, "Extended Hubbard model with the renormalized Wannier wave functions in the correlated state II: quantum critical scaling of the

wave function near the Mott-Hubbard transition," *European Physical Journal B*, vol. 74, no. 1, pp. 63–74, 2010.

[20] A. Pergament and A. Morak, "Photoinduced metal-insulator transitions: critical concentration and coherence length," *Journal of Physics A: Mathematical and General,* vol. 39, no. 17, pp. 4619–4623, 2006.

[21] M. M. Qazilbash, M. Brehm, Byung-Gyu Chae, P.-C. Ho, G. O. Andreev, Bong-Jun Kim, Sun Jin Yun, A. V. Balatsky,M. B. Maple, F. Keilmann, Hyun-Tak Kim, and D. N. Basov, "Mott transition in $VO_2$ revealed by infrared spectroscopy and nano - imaging," *Science*, vol. 318, no. 5857, pp. 1750–1753, 2007.

[22] R.-P. Blum, H. Niehus, C. Hucho, R. Fortrie, M. V. Ganduglia-Pirovano, J. Sauer, S. Shaikhutdinov, and H.-J. Freund, "Surface metal-insulator transition on a vanadium pentoxide (001) single crystal," *Physical Review Letters*, vol. 99, no. 22, Article ID 226103, 2007.

[23] M. Kang, I. Kim, S. W. Kim, J.-W. Ryu, and H. Y. Park, "Metal-insulator transition without structural phase transition in $V_2O_5$ film," *Applied Physics Letters*, vol. 98, no. 13, Article ID 131907, 2011.

[24] G. S. Nadkarni and V. S. Shirodkar, "Experiment and theory for switching in Al/$V_2O_5$/Al devices," *Thin Solid Films*, vol. 105, no. 2, pp. 115–129, 1983.

[25] E. E. Chain, "Optical properties of vanadium dioxide and vanadium pentoxide thin films," *Applied Optics*, vol. 30, pp. 2782–2787, 1991.

[26] H. Jerominek, F. Picard, and D. Vincent, "Vanadium oxide films for optical switching and detection," *Optical Engineering*, vol. 32, no. 9, pp. 2092–2099, 1993.

[27] Y. Dachuan, X. Niankan, Z. Jingyu, and Z. Xiulin, "Vanadium dioxide films with good electrical switching property," *Journal of Physics D: Applied Physics*, vol. 29, no. 4, pp. 1051–1057, 1996.

[28] P. Kiri, G. Hyett, and R. Binions, "Solid state thermochromic materials*," Advanced Materials Letters*, vol. 1, no. 2, pp. 86–105, 2010.

[29] J. Nag and R. F. Haglund Jr., "Synthesis of vanadium dioxide thin films and nanoparticles," *Journal of Physics Condensed Matter*, vol. 20, no. 26, Article ID 264016, 2008.

[30] A. L. Pergament, G. B. Stefanovich, N. A. Kuldin, and A. A. Velichko, "On the Problem of Metal-Insulator Transitions in Vanadium Oxides," *ISRN Condensed Matter Physics*, Volume 2013, Article ID 960627, 6 pages, 2013.

[31] A. Pergament, G. Stefanovich, and V. Andreev, "Comment on "Metal-insulator transition without structural phase transition in $V_2O_5$ film," [Appl. Phys. Lett. 98, 131907 (2011)]," *Applied Physics Letters*, vol. 102, Article ID176101, 1 page, 2013.

[32] C. R. Aita, "Additional Comment on "Metal-insulator transition without structural phase transition in $V_2O_5$ film," [Appl. Phys. Lett. 98, 131907 (2011)]," *Applied Physics Letters*, vol. 104, Article ID 176101, 2014.

[33] B. Pasternak, "Waves" (Translated by E. M. Kayden), 1931, Available at: http://www.friends-partners.org/friends/culture/literature/20century/pasternak/pass.html.

[34] T. Reeswinkel, D. Music, and J. M. Schneider, "Coulomb-potential-dependent decohesion of Magnéli phases," *Journal of Physics Condensed Matter*, vol. 22, no. 29, Article ID 292203, 2010.

[35] Mott insulator. Available at: http://en.wikipedia.org/wiki/Mott_insulator.

[36] J. H. de Boer and E. J. W. Verwey, "Semiconductors with partially and with completely filled 3d-lattice bands", *Proceedings of the Physical Society*, vol. 49, pp. 59-71, 1937.

[37] N. F. Mott and R. Peierls, "Discussion of the paper by de Boer and Verwey", *Proceedings of the Physical Society*, vol. 49, no. 4S, pp. 72-73, 1937.

[38] N. F. Mott, "The basis of the electron theory of metals, with special reference to the transition metals," *Proceedings of the Physical Society. Series A*, vol. 62, no. 7, pp. 416-422, 1949.

[39] M. Di Liberto, D. Malpetti, G. I. Japaridze, and C. Morais Smith, "Ultracold fermions in a one-dimensional bipartite optical lattice: Metal-insulator transitions driven by shaking," *Phys. Rev.* A 90, Article ID 023634, 2014.

[40] R. Matsunaga, N. Tsuji, H. Fujita, A. Sugioka, K. Makise, Y. Uzawa, H. Terai, Z. Wang, H. Aoki, and R. Shimano, "Light-induced collective pseudospin precession resonating with Higgs mode in a superconductor," *Science*, Vol. 345 no. 6201 pp. 1145-1149, 2014.